\documentclass[showkeys,aps,superscriptaddress,amsmath,longbibliography]{revtex4-1}
\usepackage{graphicx,txfonts}
\usepackage[colorlinks,breaklinks,linkcolor=blue,citecolor=blue,anchorcolor=blue,urlcolor=blue]{hyperref}

\begin{document}

\title{History-dependent percolation on multiplex networks}
\thanks{M.L., L.L., and Y.D. designed the research. M.L. performed the simulation and theoretical analysis. H.W. and L.L. analyzed the results on brain networks. H.W. performed the data visualization. M.L. and L.L. analyzed data with Y.D., M.-B.H., M.M. and H.E.S.'s suggestions. M.L., L.L., M.M., and H.E.S. wrote the manuscript. All the authors participated in the revisions of the manuscript.}

\author{Ming Li}
\affiliation{Department of Thermal Science and Energy Engineering, University of Science and Technology of China, Hefei 230026, P. R. China}

\author{Linyuan L\"{u}}
\email{linyuan.lv@uestc.edu.cn}
\affiliation{Institute of Fundamental and Frontier Sciences, University of Electronic Science and Technology of China, Chengdu 610054, P. R. China}
\affiliation{Alibaba Research Center for Complexity Sciences, Hangzhou Normal University, Hangzhou 310036, P. R. China}

\author{Youjin Deng}
\affiliation{Hefei National Laboratory for Physical Sciences at Microscale, Department of Modern Physics, and CAS Center for Excellence and Synergetic Innovation Center in Quantum Information and Quantum Physics, University of Science and Technology of China, Hefei 230026, P. R. China}

\author{Mao-Bin Hu}
\affiliation{Department of Thermal Science and Energy Engineering, University of Science and Technology of China, Hefei 230026, P. R. China}

\author{Hao Wang}
\affiliation{Institute of Fundamental and Frontier Sciences, University of Electronic Science and Technology of China, Chengdu 610054, P. R. China}

\author{Mat\'{u}\v{s} Medo}
\affiliation{Institute of Fundamental and Frontier Sciences, University of Electronic Science and Technology of China, Chengdu 610054, P. R. China}

\author{H. Eugene Stanley}
\affiliation{Alibaba Research Center for Complexity Sciences, Hangzhou Normal University, Hangzhou 310036, P. R. China}
\affiliation{Department of Physics and Center for Polymer Studies, Boston University, Boston, Massachusetts 02215, USA}

\date{\today}

\begin{abstract}
The structure of interconnected systems and its impact on the system dynamics is a much-studied cross-disciplinary topic. Although various critical phenomena have been found in different models, the study on the connections between different percolation transitions is still lacking. Here we propose a unified framework to study the origins of the discontinuous transitions of the percolation process on interacting networks. The model evolves in generations with the result of the present percolation depending on the previous state and thus is history-dependent. Both theoretical analysis and Monte Carlo simulations reveal that the nature of the transition remains the same at finite generations but exhibits an abrupt change for the infinite generation. We use brain functional correlation and morphological similarity data to show that our model also provides a general method to explore the network structure and can contribute to many practical applications, such as detecting the abnormal structures of human brain networks.
\end{abstract}


\maketitle

\tableofcontents

\newpage

\section{Introduction}

Our understanding of percolation properties of networks has expanded significantly in recent years. Percolation theory, a classical model in statistical physics, has been applied in a number of different network science topics, such as network structure~\cite{PhysRevE.80.036105,PhysRevLett.96.040601}, network robustness~\cite{PhysRevE.64.026118,PhysRevLett.85.5468,PhysRevLett.85.4626}, node ranking \cite{Ji2017Effective}, community detection \cite{palla2005uncovering}, as well as in studies of network dynamics, such as information spreading \cite{PhysRevE.66.016128}, and highway traffic flows \cite{Li20012015}.

Although we often assume that the underlying structure of a network is complex, the rules for the percolation process are comparatively simple. Typically, each link or node is occupied with a given probability $p$, independent of the states of other links and nodes. By contrast, real-world network processes are often ``history-dependent'': for example, the spread of a particular disease can depend on the spread of other diseases \cite{Newman2013} and it can be also influenced by the availability of immunization information~\cite{Sneppen2010}. Network topology itself can be affected by cascading failures \cite{Buldyrev2010,Gao2012} or by recovery processes \cite{majdandzic2014spontaneous} in other networks. The universality class of a percolation transition depends on the quenched disorder topology induced by the previous percolation transition \cite{PhysRevE.92.010103}.

A second complication is due to the presence of multiple interaction channels that are often involved in history-dependent processes \cite{Boccaletti20141,Gao2014}. Such systems can be naturally described in terms of multiplex networks where nodes are connected through different types of links\cite{bianconi2018multilayer}, such as the social networks with different types of interactions which can be either online or offline \cite{Szell2010PNAS,PhysRevLett.111.128701}, the multilayer transportation network with various means of vehicles \cite{Morris2012Transport}, and the brain network with both functional correlation and morphological similarity \cite{Bullmore2009Complex}. One of the typical examples of iterative interactions on multiplex networks is the interplay between the spreading of an epidemic and the information awareness that prevents its further spreading \cite{PhysRevLett.111.128701}. Another example is that of cascading failures on coupled networks of power distribution and communications \cite{Buldyrev2010,Bianconi2014}. Although these works consider a similar mechanism (\emph{i.e.}, a history-dependent process), the former features a continuous percolation transition, while the latter features a discontinuous phase transition\cite{PhysRevLett.109.248701}. We address here the question, whether there is a general model of history-dependent percolation on multiplex networks where both continuous and discontinuous percolation transitions can emerge.

To answer this question, we introduce an iterative percolation model on multiplex networks. The percolation of each generation is based on the resulting state of the previous generation, which is referred to as history-dependent percolation here. Different with the previous percolation models with some iterative processes, our work focuses also on the intermediate states of the iterative process; these intermediate states are referred to as generations here. The benefits of doing so are twofold. First, while individual intermediate generations can have their direct real counterparts, they are overlooked by focusing on the infinite (steady-state) generation. Second, by examining generations in succession, we gain understanding of the origin of the discontinuous transition in the steady state and its relation with the continuous transition.

Theoretical analysis indicates that the intermediate states of the recursive process are not cluttered, hence the percolation transition can be observed in any generations. Monte Carlo simulations on Erd\H{o}s--R\'{e}nyi (ER) networks further suggest that all these continuous transitions belong to the same universality class. Although the size of the giant cluster becomes smaller and smaller as the generations progress, endless iterations cannot completely destroy the network when it is initially dense enough. Instead, a non-vanished cluster suddenly appears above the threshold indicating a discontinuous percolation transition. Specifically, scale-free (SF) networks with exponent $2<\gamma<3$ have a vanished critical point for any finite generations, and the non-trivial critical point can suddenly emerge when the number of percolation generations diverges. With the example of human brain networks, our model shows that to find a meaningful structure (such as the abnormal structures of human brain networks), it's not always necessary to evolve the recursive process into the steady states. Our model thus provides a novel approach to analyze the network structure.

\section{Results}

\subsection{History-dependent Model}
\begin{figure}
	\centering
	\scalebox{1}[1]{\includegraphics{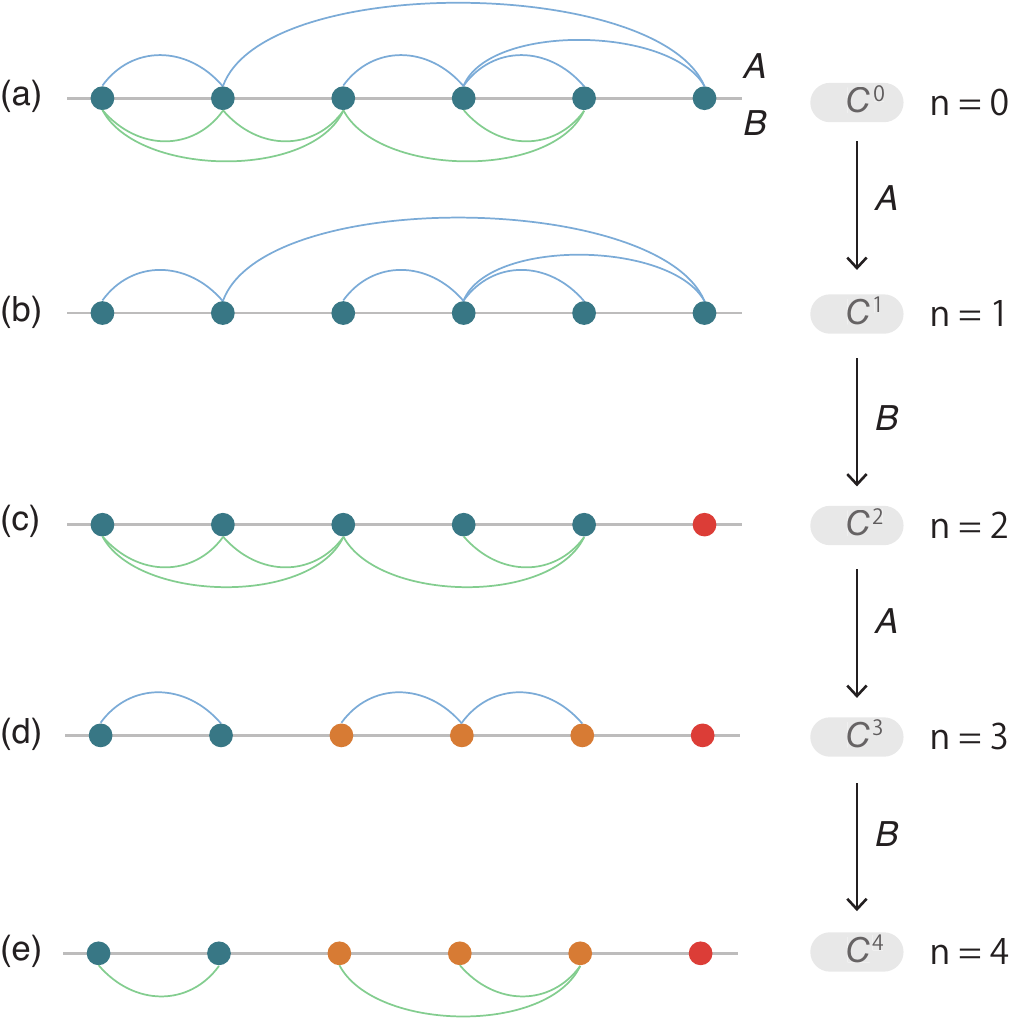}}
	\caption{Sketch of the history-dependent percolation on a small multiplex network. (a) A multiplex network with two layers $A$ and $B$ separated by the horizontal gray line. (b) Generation $n=1$. The configuration $C^1$ induced by layer $A$ is connected. (c) Generation $n=2$. The configuration induced by layer $B$ on $C^1$ has two clusters indicated by blue and red. (d) Generation $n=3$. The configuration $C^3$ induced by layer $A$ on $C^2$ has three clusters indicated by blue, orange and red, respectively. (e) Generation $n=4$. Applying layer $B$ on configuration $C^3$, no new clusters can be found; the percolation process has reached a steady state. If the system is large enough, this process can be done to any number of generations.}
	\label{fig1}
\end{figure}

An undirected multiplex network is formed by a set of $N$ nodes and multiple layers with links. Each layer is described by its adjacency matrix whose unit elements correspond to links between the corresponding nodes. For simplicity and without loss of generality, we consider here only the case of a multiplex network with two layers which we refer to as layers $A$ and $B$, respectively (see Fig.~\ref{fig1}(a) as an example). An extension to the general case with more layers is straightforward, and some discussion on this can be found in Supplementary Information (SI).

To model the history-dependent iterative process, we use the two network layers alternately to investigate the percolation process in every generation. For generation $n=1$, we use layer $A$ to check the percolation process among all the nodes. Then, nodes form a configuration $C^1$, see Fig.~\ref{fig1}(b). The percolation of generation $n=2$ is induced by layer $B$ based on the configuration $C^1$ formed in generation $n=1$. Specifically, if two nodes $i$ and $j$ who are in the same cluster in $C^1$ are connected in layer $B$, then they will be connected in this generation. Traverse all node pairs in the same clusters, we obtain a new configuration $C^2$, see Fig.~\ref{fig1}(c). Similarly, for generation $n=3$, we use layer $A$ to perform a percolation on configuration $C^2$. If two nodes $i$ and $j$ who are in the same cluster in $C^2$ are connected in layer $A$, they will be connected in this generation. Then, a new configuration $C^3$ is obtained, see Fig.~\ref{fig1}(d), and so on, to any number of generations. Note that the two layers $A$ and $B$ are used cyclically in this progressive process.

Note that once the initial configuration of the two layers is given, the constructions of clusters in all generations are deterministic. To study the percolation transition in each generation, a natural choice of the control parameter is the average degree $z$ of the initial network layers. For a network ensemble with a fixed average degree (or for a given real multiplex network), one can study the percolation transition by introducing the link occupation probability $p$ as the control parameter: the fraction $1-p$ of links in each layer are chosen at random and removed, and the remaining links are then used in the iterated percolation processes.

\subsection{Network with ER layers}
\begin{figure}
\centering
\scalebox{1}{\includegraphics{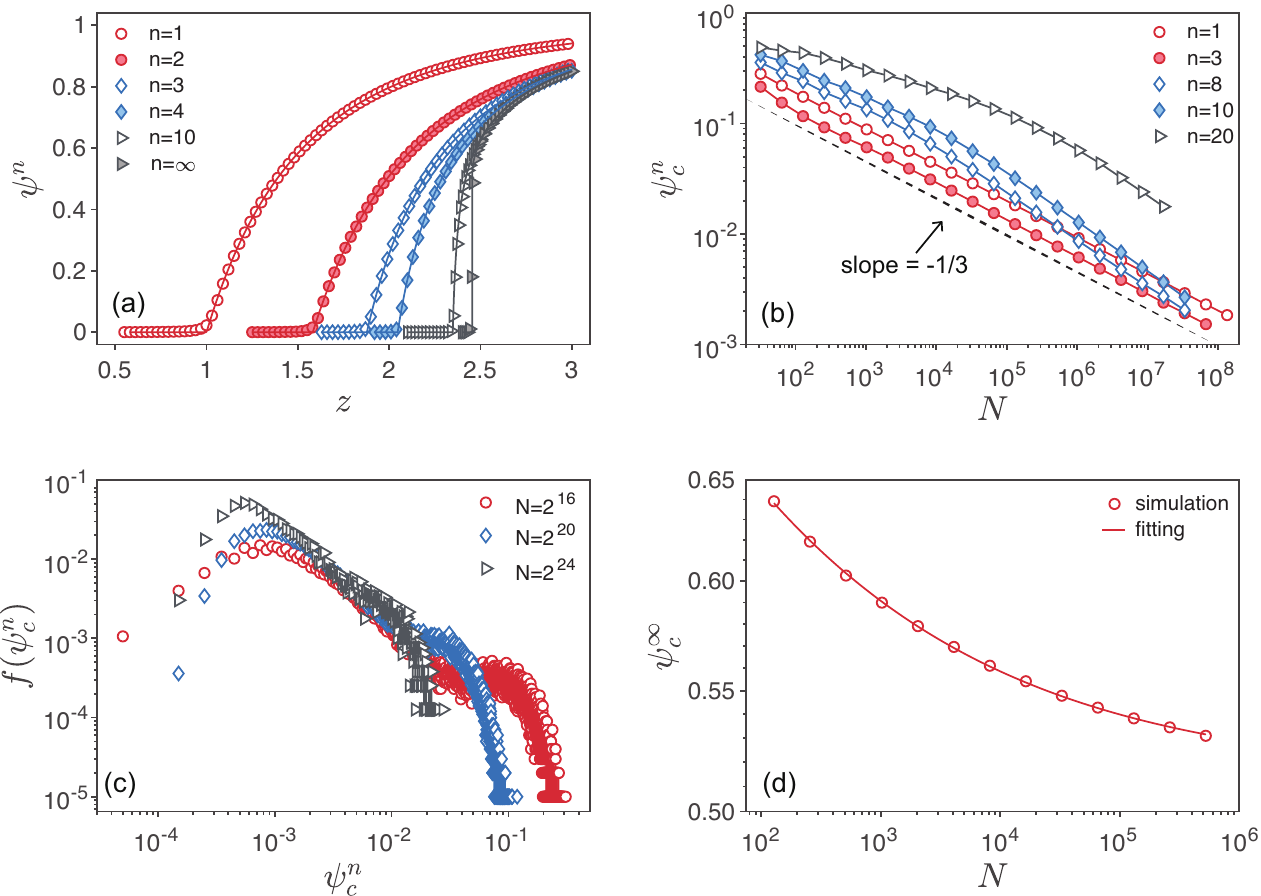}}
\caption{The simulation results for ER networks. (a) The size of the giant cluster $\psi^n$ as a function of the average degree $z$ for different generations. The solid lines are the theoretical results obtained by our method. The system size in simulations is $N=2^{16}$. (b) The order parameter at the critical point $\psi_c^n$ for finite generations as a function of the network size $N$. The simulation results are obtained by averaging over all the model realizations. (c) The distribution of the order parameters obtained in each individual realization $f(\psi_c^n)$ for generation $n=10$. (d) The order parameter at the critical point $\psi_c^\infty$ for the infinite generation as a function of the network size $N$. The simulation results are the average over the roughly $60\%$ of model realizations that percolate. The fitted curve has the form $\psi_c^{\infty}= \psi_{c0}^\infty+O(N^{-\varepsilon})$ with $\psi_{c0}^\infty=0.514\pm0.001$ and $\varepsilon=0.233\pm0.005$.}
\label{fig2}
\end{figure}

We first consider the case where the two network layers are both ER networks with the same average degree $z$, for which the model can be solved exactly (see Section Method). The results indicate that the first generation and also all finite generations of iterative percolation demonstrate a continuous percolation transition. Figure~\ref{fig2}(a) shows that the computed order parameter $\psi^n$ agrees well with the numerical simulations of the process. The theoretical solution also shows that the critical point $z_c^n$ does not diverge with the increasing generation $n$, but rather trends to a fixed value $z_c^{\infty}\approx2.455$, at which $\psi_c^\infty\approx 0.512$. This means that the percolation transition becomes discontinuous when $n\to\infty$.

To clarify the types of the percolation transition from simulations, we study the finite-size scaling of the order parameter $\psi^n_c$ at the critical point (see Section Method). Figure~\ref{fig2}(b) shows that the simulation results of the first several generations have the same scaling for large $N$, which indicates the order parameter $\psi^n_c$ will vanish when $N\to\infty$. This implies these transitions are all continuous.

\begin{figure}
	\centering
	\scalebox{1}{\includegraphics{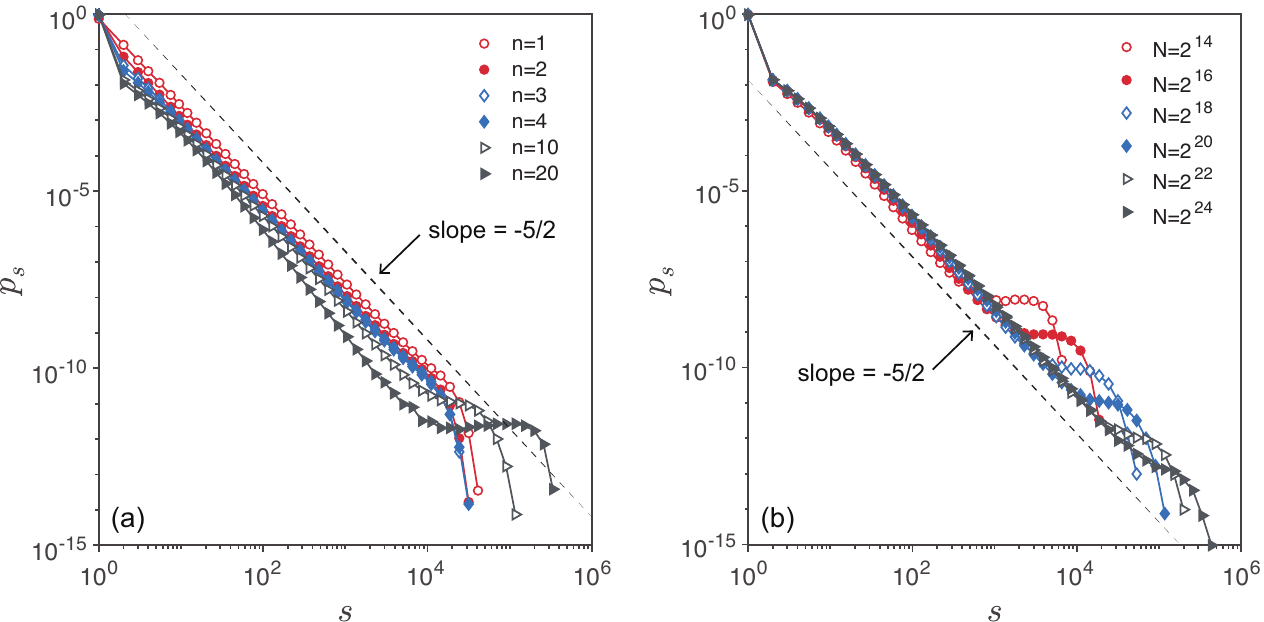}}
	\caption{The cluster size distribution $p_s$ at the critical point. (a) $p_s$ for different generations $n$; the network size is $N=2^{20}$. (b) $p_s$ of generation $n=10$ for different network sizes $N$.}
	\label{fig3}
\end{figure}

However, for a large $n$, the simulation results appear to deviate from the finite-size scaling of a continuous phase transition, see Fig.~\ref{fig2} (a). For a better understanding of this, we further show the distribution of $\psi^n_c$ obtained in each individual realization. Figure ~\ref{fig2} (c) takes generation $n=10$ as an example, see Section C of SI for other generations. The results show a heavy-tailed distribution instead of being confined to a small region as that of the classical percolation transition\cite{aharony2003introduction}. Especially when the system size is small (see the case $N=2^{16}$), the bimodal distribution as that of the discontinuous percolation transition can be found, namely, $\psi^n_c$ around zero corresponds to the non-percolating realizations and the larger values for the percolating realizations. That is why we cannot observe a finite-size scaling in Fig.~\ref{fig2} (b) for large $n$. In addition, Fig.~\ref{fig2} (c) also shows that the heavy tail will disappear with the increasing of the system size. Thus, the same scaling as that of the classical percolation is expected for very large systems.

For $n=\infty$, the distribution of $\psi^n_c$ becomes a standard bimodal distribution (see Section C of SI), which allows us to identify the non-percolating realizations and remove them from the subsequent analysis. Then, the fitted result of the percolating realizations $\psi_{c0}^{\infty}\approx 0.514$ shown in Fig.~\ref{fig2}(d) is agreement well with the theoretical analysis, indicating a discontinuous percolation transition.

Moreover, for a finite system the infinite generation just corresponds to a generation $n_c$, for which the late generations do not further alter the results. When $n\to n_c$, the system will demonstrate a much sharped percolation transition as that for the thermodynamic limit. However, $n_c$ is varied for different network realizations, and the corresponding largest clusters are thus much different. As a result, the simulation results of the largest clusters at the theoretical threshold are distributed broadly. Consequently, the heavy-tailed distribution is found (see Fig.~\ref{fig2}(c)), and the order parameters obtained by averaging over these become larger than the expectation of the finite-size scaling (see Fig.\ref{fig2} (b)). Note that $n_c$ generally increases with the system size, thus the heavy-tailed distribution shown in Fig.\ref{fig2} (c) is more obvious for smaller systems.

Since $p_c^{n-1}<p_c^{n}$, the percolation transition of generation $n$ just corresponds to a classical percolation process (with a rescaled control parameter) in the supercritical phase of generation $n-1$. This indicates that there is no essential difference between the percolation transition of two consecutive generations. As an immediate consequence, all the finite generations must belong to the same universality as the classical percolation, and the scaling behavior changes abruptly from the class shown in Fig.~\ref{fig2}(b) to that of Fig.~\ref{fig2}(d) when $n\to\infty$.

To further confirm the universality class of the history-dependent percolation model, we study the cluster size distribution $p_s$. As pointed above, the largest clusters around the critical point are distributed broadly for a large $n$. A direct result of this is increasing the probability that finding large clusters in the system. For the cluster size distribution, this results in a cocked tail before a normal exponential cutoff, and becomes more and more apparent as the increasing of generation $n$ (see Fig.\ref{fig3}(a)). However, from Fig.~\ref{fig3}(b), we can find this phenomenon weakens with the increasing of the system size $N$, so the distribution as that of the classical percolation can be expected for larger systems, \emph{i.e.}, a pow-law with an exponential cutoff (without a cocked tail). This also suggests that the universality class is the same for all finite generations of the iterated percolation process.

\subsection{Network with SF layers}
We now study the case where both layers are SF networks. In particular, we assume the power-law degree distribution~\cite{Cohen2010Complex}
\begin{equation}
\label{SF-p_k}
p_k =ck^{-\gamma},\qquad k=m, m+1, \dots, K,
\end{equation}
where $c$ is a normalization factor, and $m$ and $K$ are the lower and upper bounds of degree, respectively. If $K$ is large enough and $\gamma>1$, the normalization factor is approximately $c\approx(\gamma-1)m^{\gamma-1}$. In the simulation, networks are constructed by generating node degree values with Eq.~(\ref{SF-p_k}) and then connecting the nodes with the configuration model. Since the average degree is fixed by Eq.~(\ref{SF-p_k}), we activate a fraction $p$ of links of both network layers to control the effective mean degree and trigger the percolation process. In this section, we therefore seek the critical point in terms of critical probability $p_c$, not critical average degree $z_c$ as in the previous section.

It is known that when $\gamma>3$, such scale-free network also has a non-trivial critical point as that of ER networks~\cite{PhysRevE.64.026118}. Consequently, the results are similar to that found for ER networks. Here we focus on the case $\gamma\in(2,3)$ which is realized in many real-world networks~\cite{barabasi2016network}. Previous studies have demonstrated that the standard percolation in this case has zero critical point~\cite{PhysRevLett.85.4626,PhysRevLett.85.5468}.

\begin{figure}
\centering
\scalebox{1}{\includegraphics{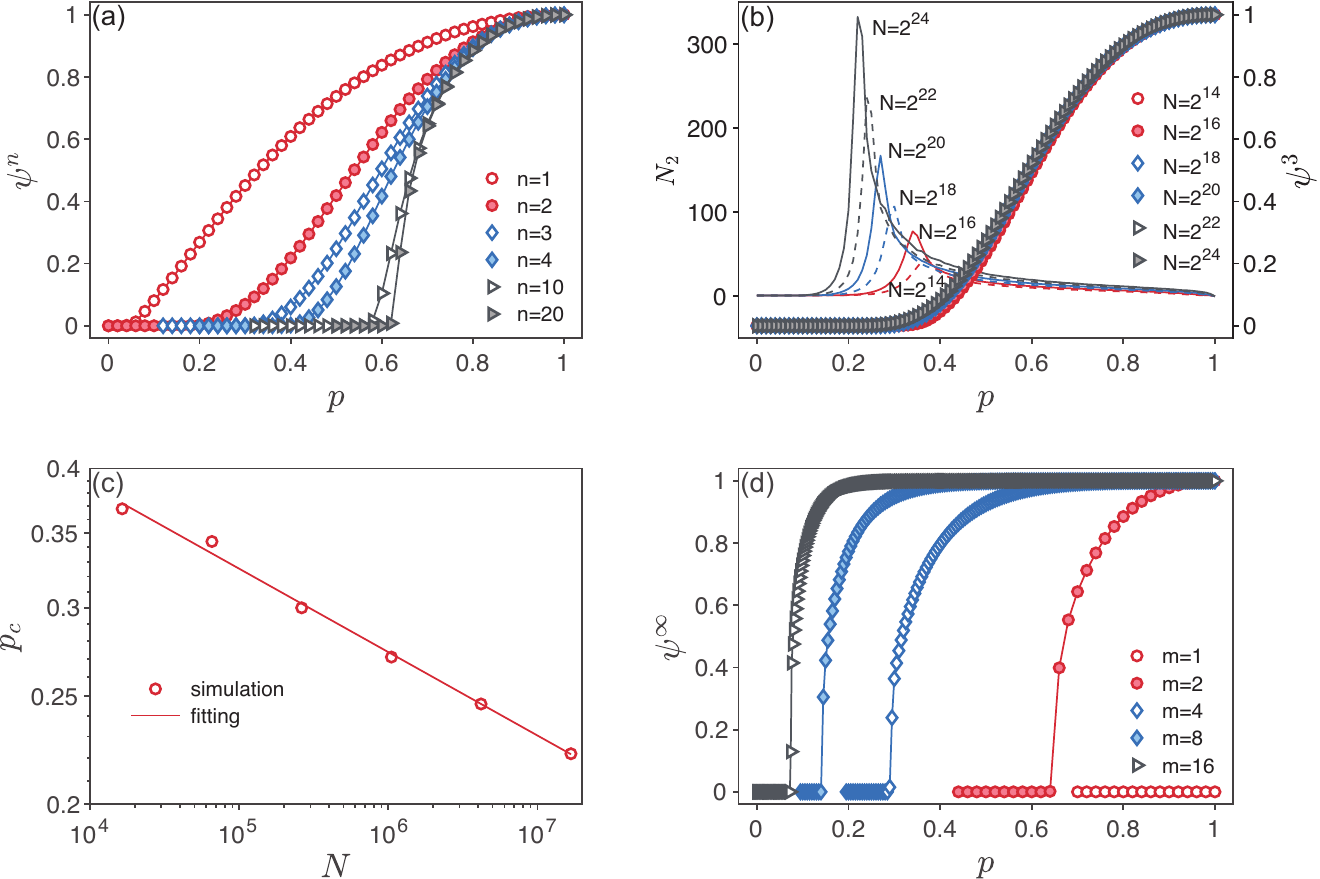}}
\caption{The simulation results for SF networks with the degree distribution given by Eq.~(\ref{SF-p_k}) where $m=2$, $K=\sqrt{N}$, and $\gamma = 2.5$. (a) The size of the giant cluster $\psi^n$ as a function of probability $p$ for different generations. The network size is $N=2^{16}$. (b) The size of the giant cluster $\psi^n$ of generation $n=3$ as a function of probability $p$ for different network sizes $N$. The solid lines are the number of nodes in the second largest cluster $N_2$. (c) The finite-size scaling of the pseudo-critical points for generation $n=3$. The fitting curve takes the form $p_c\varpropto N^{-\alpha}$ with $\alpha=0.075\pm0.003$. (d) The size of the giant cluster $\psi^n$ as a function of probability $p$ for infinite generations with different minimum degrees $m$. The network size is $N=2^{16}$.}
\label{fig4}
\end{figure}

The simulation results shown in Fig.~\ref{fig4}(a) show that the percolation transition becomes sharper and sharper as the generation increases. However, Fig.~\ref{fig4}(b) demonstrates that the pseudo-critical point indicated by the maximum of the second largest cluster decreases with the system size, which suggests a vanished critical point can also be found for infinite system. From the theoretical analysis (see Section Method and SI), we found that all the finite generations on such networks have zero critical point as the classical percolation, which is further confirmed by the finite-size scaling of pseudo-critical point shown in Fig.~\ref{fig4}(c).

We also find that to observe a non-trivial critical point on such networks, it requires endless iterations, large $m$ and broad degree distribution, otherwise the infinite generation will destroy the whole network (see Section A-C in SI for details). Generally speaking, SF networks with more connections and border degree distributions are more likely to survive in the infinite iterated processes. Figure~\ref{fig4}(d) demonstrates that when $m>1$, a discontinuous percolation transition with non-trivial critical point can also be found for SF networks.

\subsection{Real multiplex networks}
Our brain is a complex system and network neuroscience holds great promise for expanding our understanding of a healthy brain functioning, brain diseases, brain development, and brain aging\cite{sporns2005human,bassett2017network}. Simply, the brain can be modeled by a network, where the brain regions and their connections constitute the set of nodes and the set of links, respectively. Here, we consider the human brain networks which are constructed using the high-resolution brain atlas with $1024$ Regions-of-Interest (ROIs) \cite{Zalesky2010Whole}, \emph{i.e.}, the network thus has $1024$ nodes. This multiplex network has two layers, which capture functional correlations and morphological similarity of the human brain based on ROIs, respectively. For the functional brain layer, the mean time series is extracted for each ROI by averaging the time series of all voxels (small measured volumes in three-dimensional space) within it. Then, we calculate the Pearson correlation for each pair of ROIs and generated a $1024\times1024$ correlation matrix. For the morphological brain layer, we estimate the interregional similarity in the distribution of regional gray matter volume in terms of the Kullback-Leibler divergence measure~\cite{Hao2016Single}. By this construction, both layers are represented by weighted complete networks. We tune the resulting networks by choosing the average degree in each layer, $z$, and thus obtain the corresponding unweighted networks where only links with the highest weights are kept. See the last section of SI for a detailed description of the data and the MRI data preprocessing strategy.

\begin{figure}
\centering
\scalebox{1}{\includegraphics{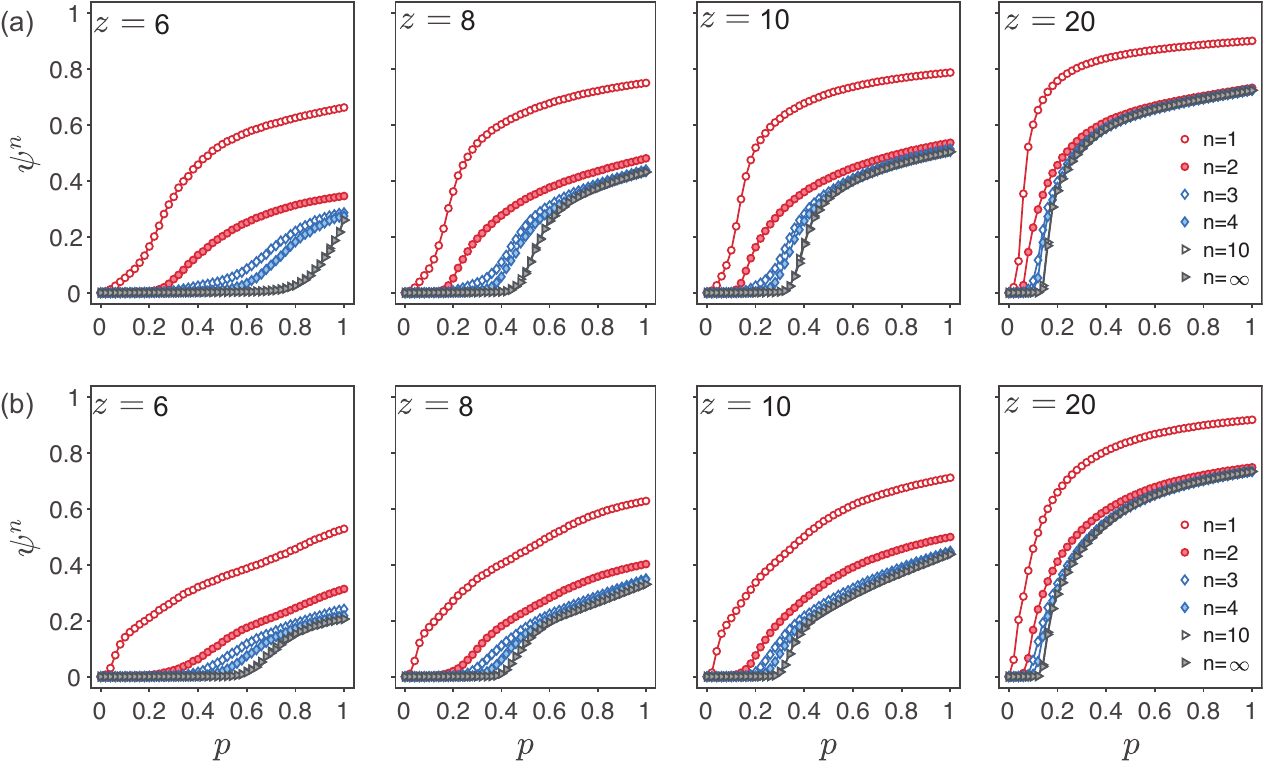}}
\caption{History-dependent percolation on human functional-morphological brain networks for different average degree values ($z$); $p$ is the link occupation probability; $\psi^n$ is the size of the giant cluster given the percolation generation $n$. (a) The HC participant's bilayer brain network. (b) The MDD participant's bilayer brain network.}
\label{fig5}
\end{figure}

In Fig.~\ref{fig5}, we compare the results of our percolation model on two different human functional-morphological brain networks, one from a major depressive disorder (MDD) participant and a healthy control (HC) participant. For the MDD data, both the critical point and the giant cluster for a given $p$ are smaller than that of HC, indicating that the brain network of the MDD participant is more vulnerable. For the infinite generation of the model, the pattern of the remaining nodes and links (the giant cluster) of the MDD data is sparser and more dispersed. These nodes mainly locate in frontal, parietal, and occipital lobes, and don’t form obvious community structure, whereas, for the HC participant, these nodes are close together and mainly locate in the highly myelinated brain regions (i.e., the motor-somatosensory strip in the central sulcus, the visual cortex in the occipital lobe). This suggests that for the HC participant, the giant cluster  identified by a finite-generation percolation process may reflect the biological meanings. See Fig.7 in SI for details.

\begin{figure}
\centering
\scalebox{1}{\includegraphics{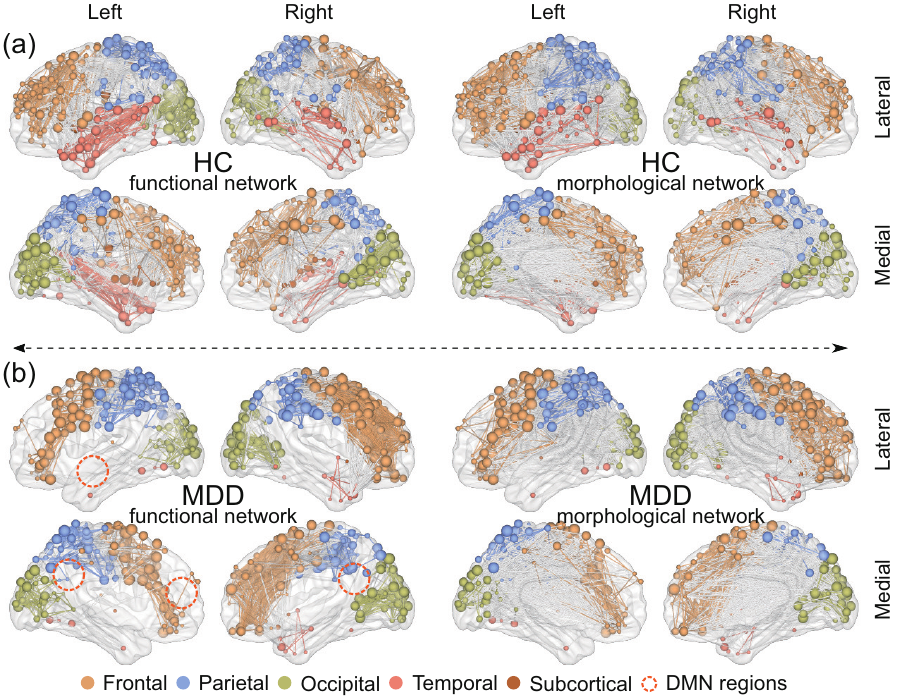}}
\caption{The visualization of human brain bilayer networks at degree $z=20$ and $1024$ parcellation templates when $p$ is slightly larger than the corresponding $p_c$. (a) The HC's bilayer brain network. (b) The MDD's bilayer brain network. We show the lateral and medial brain of each hemisphere. The visualization is done with BrainNet viewer~\cite{Xia_2013}.}
\label{fig6}
\end{figure}

When we increase the average degree to $z = 20$, both brain networks become robust, and the percolation transition of infinite generation becomes sharper (see Fig.~\ref{fig5}). As shown in Fig.~\ref{fig6}, for the HC participant, the giant cluster of the infinite generation diffuses across frontal, parietal, occipital, temporal, and subcortical lobes. Compared with HC data, the MDD network shows an apparent deficiency of nodes in frontal, occipital, and temporal lobes, some of which belong to the default mode network (DMN) \cite{Raichle_2001}. These DMN regions play an important role in our advanced cognitive abilities, such as executive control, visual and auditory sense and some pieces of evidence support the decreased functional connectivity \cite{Veer,Zhang} and frontal cortical thinning \cite{Bos} in these regions for MDD participants, see Fig.~\ref{fig6}. Therefore, our model opens a new avenue toward detecting the abnormal nodes or components between the healthy and the diseases participants from a comprehensive functional-morphological perspective, which can help us detect the potential connectome-based MRI biomarker and gain new insights on the mechanism for some brain disorders. Importantly, these findings are beyond the single modal imaging/layer and traditional percolation, thus provide the novel understanding and convergent results for MDD. Furthermore, our model is easy to expand to the networks with three or more layers and could be used to investigate the similarities, differences and comprehensive understanding of various brain disorders. Note that the current study is a single case validation. Besides, our brain is highly personalized and a large sample size is needed to infer robust conclusions \cite{Termenon,Turner}. Further study can utilize large sample size or different brain disorders to verify our model.

We also applied our model to a bilayer social network composed of users who are active on both Twitter and FriendFeed~\cite{Magnani2011The}. Among the $150,684$ common users of the two networks, there are $8,308,326$ and $5,270,665$ links in the Twitter and FriendFeed layer, respectively. The results can be found in Fig.6 in SI. We find that the percolation transition occurs at much lower activation probability $p$ than that of the brain network data. This is a direct consequence of the average degree in the social network ($z\approx110$ and $70$ in the two network layers, respectively) being substantially higher than in the brain network ($z=6,8,10,20$). The Twitter-FriendFeed network reaches the steady state after $4$ generations, and the discontinuous percolation transition is also absent. This can be due to many links (about $37\%$ of the total number) that occur in both network layers which renders subsequent process generations equivalent to the first two or three generations of the case discussed in Fig.~\ref{fig4}.

These results suggest that to observe non-trivial behaviors in iterative percolation, one needs to study multiplex networks with limited layer overlap. To fully understand the relation between the layer overlap and the dynamics of the iterative percolation remains a future challenge. The human brain network reaches the steady state after $10$ generations. At this point, the percolation transition is more abrupt than that for smaller iteration values. With respect to the small size of the studied system ($1024$ nodes), the possibility that a larger brain network (achievable with a higher imaging resolution) would display yet more abrupt transition, and thus suggest a discontinuous percolation transition in the thermodynamic limit, remains open. However, whether the percolation transition is discontinuous or not, the process provides a series of method to analyze network structures.

\section{Discussion}
In summary, we introduce a history-dependent percolation model to study the critical behavior of the percolation transition on multiplex networks. The percolation process is ran iteratively on different layers. Instead of focusing solely on the steady state, we pay more attention to finite generations, each of which can be considered as an independent model. For example, $n=1$ corresponds to the common model of information or disease spreading on monopartite networks; $n=2$ can be used to describe the interplay between the spreading of disease and immunity information \cite{PhysRevLett.111.128701}; $n=3$ can be used to model the information spreading on multiplex social networks where users communicate through multiple channels. The general example of multilayer propagation is the one where online communication influences user offline behavior and then couples back to online communication. A typical example of $n=\infty$ is the study of cascading failures in coupled networks \cite{Buldyrev2010}. In this sense, our model provides a unified framework to study the percolation process on multiplex networks.

We investigate both ER networks and SF networks with power-law exponent $2<\gamma<3$. The results reveal that the intermediate state of the recursive process can be also used to uncover meaningful structures, and therefore the percolation transition should be defined and studied in each generation. For any finite generations, the percolation transition on random networks show a continuous transition and belong to the same universality class, while scale-free networks have the critical point trending to zero as the network size grows. When $n=\infty$, a discontinuous transition exists for both networks \cite{Buldyrev2010}. In essence, this is due to that the percolation transition of generation $n$ emerges from the supercritical phase of the percolation of generation $n-1$. As a result, it inherently cannot generate a new universality class\cite{PhysRevE.92.010103}. However, when $n$ diverges, $n-1$ is also divergent, this relation of the critical state is broken. Then, the new phenomenon, \emph{i.e.}, discontinuous transition, emerges.

Furthermore, on one hand, our result indicates that the continuous transition found in real systems could be the result of a combination of many sequential processes, see an example in Ref.~\cite{PhysRevLett.111.128701}. On the other hand, to observe the abrupt percolation transitions in real multiplex networks, for which the generation cannot go to infinite actually, the model may need to be extended by, for example, considering the space embedding of network layers \cite{bashan2013extreme}, modeling relative importance of inter- and intra-layer connections \cite{Radicchi2015Percolation}, or introducing cores of ``high quality" edges~\cite{radicchi2013abrupt}. Our model can be easily extended to a general case with more layers. This would make it possible to use the generalized model to analyze the percolation transition on a temporal network comprising several layers corresponding to different times points \cite{Holme2012Temporal}.

Beyond the theoretical analysis, we showed that the outcomes of the iterative percolation process can be used to characterize real networks whose percolation properties differ markedly between systems (such as the used brain scan and social network data) as well as between various samples of networks from the same class (such as the brain scan data of a healthy participant and a participant with a mental disorder). The proposed model can thus become a useful tool for evaluating and, more importantly, comparing structural properties of multiplex networks. The model represents an important step towards understanding the history-dependent dynamic processes on multiplex networks, and may prove useful in important practical applications like link prediction \cite{L2011Link}, vital node identification \cite{L2016Vital} and community detection \cite{PhysRevX.7.031056} in multiplex networks.

\section{Method}
\label{sec:method}

\subsection{Mean-field theory analysis}
To obtain the exact analytical solution of the critical point, we consider the giant cluster in the infinite system~\cite{Buldyrev2010}. We introduce first the function $\mathfrak{F}(x)$ which returns the size of the giant cluster of a network ensemble with given degree distribution, when a fraction $x$ of nodes are chosen at random and used to construct the giant cluster. Note that the fraction obtained by function $\mathfrak{F}(x)$ is with respect to the actually used nodes. The size of the giant cluster with respect to the original network is thus $x\mathfrak{F}(x)$.

Leaving the specific form of $\mathfrak{F}(x)$ aside, which may be a group of equations or a dataset obtained by Monte Carlo simulations, we now lay out the general analytical framework for the history-dependent percolation process. Since the network's layers are in general different, we assume that layers $A$ and $B$ have functions $\mathfrak{F}_A(x)$ and $\mathfrak{F}_B(x)$, respectively. In addition, we label the size of the giant cluster in generation $n$ as $\psi^n$, and the fraction of the nodes that can be used to construct the giant cluster in generation $n$ as $S^{n-1}$. The function $\mathfrak{F}(x)$ allows us to write
\begin{equation}
\psi^n=S^{n-1}\mathfrak{F}(S^{n-1}), \label{gnpsi}
\end{equation}
where $\mathfrak{F}(x)$ is $\mathfrak{F}_A(x)$ and $\mathfrak{F}_B(x)$ for odd and even generations, respectively. All we need to do now is to find $S^{n-1}$ for each generation $n$.

For an infinite system, the giant cluster of generation $n$ can only emerge from the giant cluster of generation $n-1$. So, leveraging this recursive relationship, the fraction of nodes that can be used to construct the giant cluster $S^{n}$ for an odd $n$ satisfies
\begin{equation}
S^{n} = S^0\mathfrak{F}_A(S^{n-1}),     \label{gns1}
\end{equation}
and for an even $n$,
\begin{equation}
S^{n} = S^0\mathfrak{F}_B(S^{n-1}).     \label{gns2}
\end{equation}
In addition, if one removes a fraction $1-p$ of nodes in the initial configuration to trigger the iterated percolation, then $S^0=p$. If the removal is for links, such as the ones used in Figs.~\ref{fig4} and \ref{fig5}, the function $\mathfrak{F}(x)$ for the diluted network should be replaced with $\mathfrak{F}(px)$ since the degree distribution has changed~\cite{PhysRevE.66.016128}.

If functions $\mathfrak{F}_{A,B}(x)$ are known, Eqs. (\ref{gnpsi})-(\ref{gns2}) can be used to calculate the size of the giant cluster for any generation and also the critical point. For ER networks $\mathfrak{F}(x)$ can be obtained by solving a self-consistent equation, however, there is no closed form for SF networks and the theoretical analysis can be only done around the critical point, see Section A of SI for details.

If function $\mathfrak{F}(x)$ has a non-trivial critical point $x_c$ below which $\mathfrak{F}(x)=0$ and $\mathfrak{F}(x_c)=0$, such as ER networks. Equations (\ref{gns1}) and (\ref{gns2}) shows that the critical point of generation $n$ corresponds to a non-zero $S^{n-1}$, that is the supercritical state of generation $n-1$. This indicates that there is no essential difference between the percolation transition of the two generations, and consequently the first generation and also all finite generations of iterative percolation demonstrate a continuous percolation transition. Figure~\ref{fig2}(a) takes ER networks as example to shows that the resulting $\psi^n$ agrees well with numerical simulations of the process.

The recursive relations Eqs.(\ref{gns1}) and (\ref{gns2}) have their fixed point $S=\mathfrak{F}(S)$, which corresponds to the infinite generation, and allows us to find the fixed point for ER layers $z_c^{\infty}\approx2.455$ and $S_c=\left(1+\sqrt{1-z_c^\infty/2}\right)/2\approx 0.715$, see Section A-C of SI for details. The critical order parameter that corresponds to the found $S_c$ is $\psi_c^\infty=(S_c)^2\approx 0.512$, which also agrees well with numerical simulations shown in Fig.\ref{fig2}(d). The giant cluster size thus undergoes a discontinuous phase transition at $z_c^{\infty}$.

For SF networks, the recursive relations Eqs.~(\ref{gns1}) and (\ref{gns2}) can also be used to obtain the critical point of each generation. Since $\mathfrak{F}(px)$ has a zero critical point for $2<\gamma<3$, we can find that all the finite generations give a vanished critical point $p_c=0$ by using Eqs.~(\ref{gns1}) and (\ref{gns2}), recursively (see Section A-B of SI for details). By examining the fixed point of Eqs.~(\ref{gns1}) and (\ref{gns2}), we can also find the infinite generation demonstrates a discontinuous percolation transition, see Section A-C of SI for details.

\subsection{The order parameter at the critical point}
As a topological phase transition, there is no free energy can be used to determine the type of the percolation transition. Therefore, the one with a step-like changing of the order parameter at the critical point is often treated as a discontinuous percolation transition \cite{Buldyrev2010,Boccaletti20141,PhysRevLett.96.040601,PhysRevLett.109.248701,Radicchi2015Percolation,radicchi2013abrupt}, and the continuous percolation transition is for that with continuous changing at the critical point. That is to say, the discontinuous percolation transition has a non-zero order parameter at the critical point, and a zero order parameter can be found for continuous percolation transition.

However, due to finite sizes of the systems used in the simulation, both the two types of percolation transitions could give a non-zero order parameter at the critical point. In this way, we use the finite-size scaling to check the types of percolation transition in simulations. For a continuous transition, the order parameter $\psi_c$ must decrease with increasing system size and becomes zero for an infinite system, which corresponds to $\psi_c\propto N^{-\epsilon}$ where $\epsilon=1/3$ for random networks~\cite{aharony2003introduction}. For a discontinuous transition, the order parameter $\psi_c$ takes a bimodal distribution, the values around zero correspond to the non-percolating realizations and the larger values for the percolating realizations. Excluding the non-percolating realizations, the finite-size scaling thus takes the form $\psi_c\sim \psi_{c0}+O(N^{-\varepsilon})$, where $\psi_{c0}$ is the order parameter at the critical point for an infinite system. Therefore, Figs.~\ref{fig2} (b)-(d) just indicate that all the finite generations take the continuous percolation transition, and the infinite generation takes the discontinuous percolation transition. However, due to finite-size effects, larger systems are needed to observe a better finite-size scaling.

\begin{acknowledgments}
The research was partially supported by the National Natural Science Foundation of China under grant Nos.11622538, 61773148, 61673150, 11625522 and 11672289. L.L. also acknowledges Zhejiang Provincial Natural Science Foundation of China under grant No.LR16A05000, and the Science Strength Promotion Programme of UESTC. Y.D. also acknowledges the Ministry of Science and Technology of China under grant No.2016YFA0301604.
\end{acknowledgments}

\bibliography{ref}

\newpage

\appendix

\section{Theory}

\subsection{General formalism}
We introduce the function $\mathfrak{F}(x)$ to represent the size of the giant cluster of a network ensemble with a given degree distribution, when a fraction $x$ of nodes are chosen at random and used to construct the giant cluster. Note that the fraction obtained by function $\mathfrak{F}(x)$ is with respect to the number of actually used nodes (fraction $x$ of all nodes). The size of the giant cluster with respect to the original network is thus $x\mathfrak{F}(x)$.

For our model, different layers can have different degree distributions, the above-described function then differs from one layer to another, and we can label it $\mathfrak{F}_{i}(x)$ for layer $i$. Assuming the fraction of the nodes that can be used to construct the giant cluster in generation $n$ is $S^{n-1}$, then the function $\mathfrak{F}(x)$ allows us to write the size of the giant cluster in generation $n$ as
\begin{equation}
\psi^n=S^{n-1}\mathfrak{F}_{l_n}(S^{n-1}),  \label{psi}
\end{equation}
where $l_n$ is the layer used in generation $n$.

In the infinite system, a giant cluster can only emerge from the giant cluster of the prior generation. Consequently, there is a recursive relation between the fraction $S^n$ of two successive generations, which can also be expressed by function $\mathfrak{F}(x)$,
\begin{equation}
S^{n-1} = S^0\prod_{i=(n-M)H(n-M)+1}^{n-1}\mathfrak{F}_{l_i}(S^{i-1}),~~~~n\geq 2.    \label{sn1}
\end{equation}
Here, $M$ is the number of layers, and $H(x)$ is the Heaviside step function, \emph{i.e.},
\begin{equation}
H(x)=\left\{ \begin{aligned}
               1, & ~~~~x\geq 0, \\
               0, & ~~~~x<0.
             \end{aligned}\right.
\end{equation}
Next, we take $M=2, 3$ as examples to provide an interpretation of Eq.~(\ref{sn1}).

\subsubsection{$M=2$}
The case $M=2$ is the one shown in the main text, so we use the same notation, \emph{i.e.}, the two layers are labeled $A$ and $B$. In generation $n=1$, all nodes can be used to construct the giant cluster, \emph{i.e.}, $S^0=1$. The giant cluster size $\psi^1$ can be thus expressed as $\psi^1=S^0\mathfrak{F}_A(S^0)$. The giant cluster of generation $2$ emerges from the giant cluster of generation $n=1$. The fraction of nodes $S^1=\psi^1$ is therefore the starting point for generation $n=2$ and we can write $\psi^2=S^1\mathfrak{F}_B(S^1)$.

For generation $n=3$, some nodes in the giant cluster of generation $n=1$ can no longer be used to construct the giant cluster, and the fraction is $\psi^1-\psi^2=S^1[1-\mathfrak{F}_B(S^1)]$. Removing these nodes from the giant cluster (fraction $S^1=\psi^1$) is equivalent to removing the same fraction of nodes from $S^0$, since the other nodes (fraction $1-S^1$) that are not used to construct the giant cluster of generation $n=1$ do not belong to $S^1$ and $\psi^2$. So the total fraction of nodes can not be used in generation $n=3$ is $S^0[1-\mathfrak{F}_B(S^1)]$. In this way, generation $n=3$ is equivalent to randomly using a fraction $S^2=S^0\mathfrak{F}_B(S^1)$ of nodes to construct the giant cluster. It is easy to see that the general formula for the fraction of nodes that can be used to construct the giant cluster $S^{n}$ for an odd $n$ is
\begin{equation}
S^{n} = S^0\mathfrak{F}_A(S^{n-1}),     \label{gns1s}
\end{equation}
and for an even $n$,
\begin{equation}
S^{n} = S^0\mathfrak{F}_B(S^{n-1}).     \label{gns2s}
\end{equation}
These two equations are clearly a particular form of Eq.~(\ref{sn1}).

\subsubsection{$M=3$}
We further take $M=3$ as another example. In generation $n=1$, all nodes can be used to construct the giant cluster, \emph{i.e.}, $S^0=1$. The giant cluster size $\psi^1$ can be thus expressed as $\psi^1=S^0\mathfrak{F}_1(S^0)$. The fraction of nodes $S^1=\psi^1=S^0\mathfrak{F}_1(S^0)$ is therefore the starting point for generation $n=2$ and we can write $\psi^2=S^1\mathfrak{F}_2(S^1)$. Similarly, for generation $n=3$, the fraction $S^0-\psi^2$ of nodes cannot be used to construct the giant cluster. That is $S^2=\psi^2=S^1\mathfrak{F}_2(S^1)=S^0\mathfrak{F}_1(S^0)\mathfrak{F}_2(S^1)$, and $\psi^3=S^2\mathfrak{F}_3(S^2)$.

For generation $n=4$, we will reuse layer $1$ to construct the giant cluster, however, some nodes in the giant cluster of generation $n=1$ can no longer be used to construct the giant cluster, and the fraction is $\psi^1-\psi^3=S^1[1-\mathfrak{F}_2(S^1)\mathfrak{F}_3(S^2)]$. Removing these nodes from the giant cluster (fraction $S^1=\psi^1$) is equivalent to removing the same fraction of nodes from $S^0$, since the other nodes (fraction $1-S^1$) that are not used to construct the giant cluster of generation $n=1$ do not belong to $S^1$ and $\psi^3$. So the total fraction of nodes can not be used in generation $n=4$ is $S^0[1-\mathfrak{F}_2(S^1)\mathfrak{F}_3(S^2)]$. In this way, generation $n=4$ is equivalent to randomly using a fraction $S^3=S^0\mathfrak{F}_2(S^1)\mathfrak{F}_3(S^2)$ of nodes to construct the giant cluster. By analogy, the fraction of nodes that can be used to construct the giant cluster can be represented as
\begin{eqnarray}
S^1 &=& S^0\mathfrak{F}_1(S^0), \\
S^2 &=& S^0\mathfrak{F}_1(S^0)\mathfrak{F}_2(S^1), \\
S^3 &=& S^0\mathfrak{F}_2(S^1)\mathfrak{F}_3(S^2), \\
S^4 &=& S^0\mathfrak{F}_3(S^2)\mathfrak{F}_1(S^3), \\
S^5 &=& S^0\mathfrak{F}_1(S^3)\mathfrak{F}_2(S^4), \\
    &\vdots&  \nonumber \\
S^{n} &=& S^0\mathfrak{F}_{l_{n-2}}(S^{n-2})\mathfrak{F}_{l_{n-1}}(S^{n-1}).
\end{eqnarray}
A more general expression of these equations is just Eq.~(\ref{sn1}). In addition, if one removes a fraction $1-p$ of nodes in the initial configuration to trigger the iterated percolation, then $S^0=p$. If the removal is for links, the function $\mathfrak{F}(x)$ for the reduced network should be replaced with $\mathfrak{F}(px)$.

\begin{figure}
\centering
\scalebox{1.2}{\includegraphics{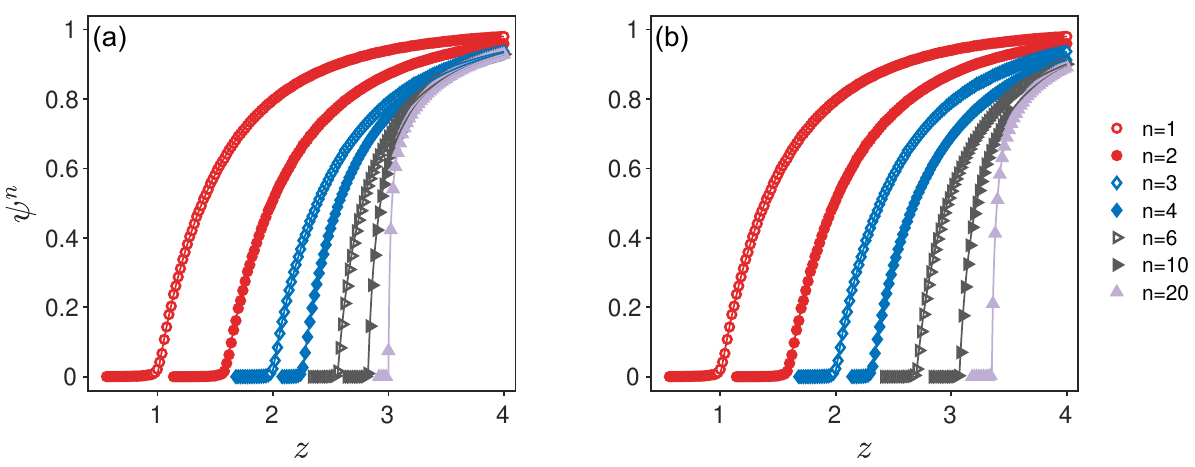}}
\caption{The size of the giant cluster $\psi^n$ as a function of the average degree $z$ for different generations $n$. In the simulation, the system size is $N=2^{16}$, and all layers are ER networks with the same average degree $z$. The solid lines are the theoretical results obtained by Eqs.~(\ref{psi}) and (\ref{sn1}), numerically. (a) $M=3$. (b) $M=4$.}
\label{fig1s}
\end{figure}

If functions $\mathfrak{F}(x)$ of all layers are known, we can get the theoretical results by Eqs.~(\ref{psi}) and (\ref{sn1}), analytically or numerically. In Fig.~\ref{fig1s}, we give the simulation results for $M=2$ and $3$, which are in agreement with the theory well.

\subsection{The critical point}
The critical point of this system is determined by function $\mathfrak{F}(x)$ in Eq.~(\ref{psi}). Assuming function $\mathfrak{F}(x)$ has a critical point $x_c$ below which $\mathfrak{F}(x)=0$. Thus, substituting $S^{n-1}=x_c$ into Eqs.~(\ref{psi}) and (\ref{sn1}), we can obtain the critical point of for any generation. Next, we take the case used in the main text to show the details, that is a two-layer network with the same degree distribution in each layer.

\subsubsection{Two-layer Erd\H{o}s--R\'{e}nyi (ER) networks}

\begin{figure}
	\centering
	\scalebox{1.3}{\includegraphics{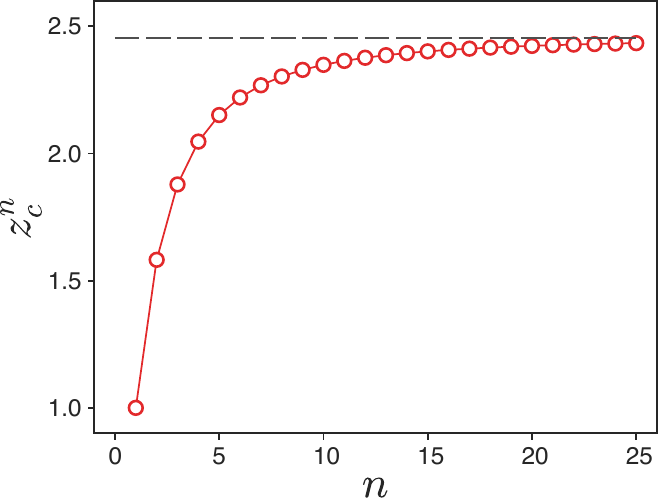}}
	\caption{The critical point $z_c^n$ of ER networks obtained by our method for different generations $n$. The dotted line is the critical point for the infinite generation, $z_c^{\infty}\approx2.455$.}
	\label{fig2s}
\end{figure}

For ER networks, function $\mathfrak{F}(x)$ satisfies the self-consistent equation
\begin{equation}
\mathfrak{F}(x)=1-e^{-zx\mathfrak{F}(x)},   \label{serp}
\end{equation}
with the critical point $x_c=1/z$ below which $\mathfrak{F}(x)=0$. According to Eq.~(\ref{psi}), the critical point of generation $n$ corresponds to $S^{n-1}=x_c=1/z$. For $S^0=1$ and $M=2$, the iterative Eq. (\ref{sn1}) simplifies to the form $S^n=\mathfrak{F}(S^{n-1})$. The critical point of generation $n$ can thus be further developed as $S^{n-2} = \mathfrak{F}^{-1}(S^{n-1})=\mathfrak{F}^{-1}(1/z)$, where $\mathfrak{F}^{-1}(x)$ is the inverse function of $\mathfrak{F}(x)$ which reads
\begin{equation}
\mathfrak{F}^{-1}(x)=-\frac{\ln(1-x)}{zx}       \label{f1}
\end{equation}
and one can easily verify that $\mathfrak{F}[\mathfrak{F}^{-1}(x)]=x$. One can continue the development with $S^{n-3}$, $S^{n-4}$, and so on until
\begin{equation}
\underbrace{\mathfrak{F}^{-1}\left(\cdots \mathfrak{F}^{-1}\left(\mathfrak{F}^{-1}(1/z_c^n)\right)\right)}_{\text{$n-1$ times}}= S^0 =1.    \label{pcn}
\end{equation}
For any generation $n$, we can use this equation to get the critical point $z_c^n$, numerically. The theoretical solution also shows that the critical point $z_c^n$ does not diverge with the increasing of generation $n$), but rather trends to a fixed value $z_c^{\infty}\approx2.455$ (see Fig.\ref{fig2s}).

Furthermore, one also can note that $\mathfrak{F}(x)\to 0$ when $x\to x_c^+$, indicating that the first generation and consequently also all finite generations of iterative percolation demonstrate a continuous percolation transition, \emph{i.e.}, $\psi_c^n=0$.

\subsubsection{Two-layer scale-free (SF) networks}
For this case, we assume that both the two layers have the same power-law degree distribution
\begin{equation}
p_k =ck^{-\gamma},\qquad k=m, m+1, \dots, K, \label{SF-p_ks}
\end{equation}
where $c$ is the normalization factor, and $m$ and $K$ are the lower and upper bounds of degree, respectively. If $K$ is large enough and $\gamma>1$, the normalization factor is approximately $c\approx(\gamma-1)m^{\gamma-1}$.

Since the average degree is fixed by Eq.~(\ref{SF-p_ks}), we randomly remove a fraction $1-p$ of links of both network layers to control the effective mean degree. Then, the function $\mathfrak{F}(px)$ for this system can be expressed by the generating functions
\begin{eqnarray}
R &=& 1-G_1(1-pRx),  \label{sf1} \\
\mathfrak{F}(px)&=&1-G_0(1-pRx),  \label{sf2}
\end{eqnarray}
where $R$ is an auxiliary variable, $G_0(x)=\sum_k p_kx^k$ and $G_1(x)=\sum_k p_kkx^{k-1}/z$ are the generating functions of the degree and excess-degree distributions, respectively. From Eqs.~(\ref{sf1}) and (\ref{sf2}), the critical point of $\mathfrak{F}(px)$ can be obtained
\begin{equation}
p_c=\frac{1}{G_1^\prime(1)},
\end{equation}
below which $\mathfrak{F}(px)=0$. Note that for $m=1$, this formula gives a percolation threshold $p_c$ larger than $1$ when $\lambda>3.47875\ldots$, indicating there is no percolation transition. This is due to the absent of the spanning cluster in such SF networks [1].

For $\gamma\in(2,3)$ which is realized in many real-world networks, $G_1^\prime(1)\to\infty$ with the increasing system size. So function $\mathfrak{F}(px)$ gives a vanished critical point $p_c=0$ with $\mathfrak{F}(p_cx)=0$. According to Eq.~(\ref{psi}), the critical point of generation $n$ corresponds to $S^{n-1}=0$. Then, using the iterative relation $S^n=\mathfrak{F}(pS^{n-1})$, it is clear that for any finite generation, the system gives a vanished critical point.

In addition, when $\gamma>3$, function $\mathfrak{F}(px)$ has a non-trivial critical point as that of ER networks. Consequently, the results are similar to that we introduced above for ER layers, \emph{i.e.}, there is a non-trivial percolation transition for any finite generation. Note that due to the strong heterogeneity of SF networks, some special critical exponents dependent on the exponent $\gamma$ can be found [2,3].

\subsection{Infinite generation}
For infinite iterations, layers are favored rather than generations. For convenience, we use subscript to differentiate layers instead of generations, such as $\psi_i$ for the size of the giant cluster of layer $i$, and $S_i$ for the fraction of nodes can be used to construct the giant cluster of layer $i$. Based on these notations, Eq.(\ref{sn1}) can be rewritten as
\begin{equation}
S_i = S^0\prod_{j\neq i}^M\mathfrak{F}_j(S_j).    \label{si}
\end{equation}
Substituting this relation into Eq.~(\ref{psi}), we have
\begin{equation}
\psi_i=S_i\mathfrak{F}_i(S_i)=S^0\prod_{j=1}^M\mathfrak{F}_j(S_j) \equiv\psi.    \label{psii}
\end{equation}
This means that all the giant clusters (in different layers) have the same size. According to Eq.~(\ref{psii}), the case of the infinite generation can be solved by the following equations
\begin{eqnarray}
\psi &=& S^0\prod_{i=1}^M\mathfrak{F}_i(S_i),   \label{psia} \\
S_i  &=& \frac{\psi}{\mathfrak{F}_i(S_i)},~~~~i=1, 2, \ldots, M. \label{sia}
\end{eqnarray}
This form recovers the finding of the percolation on tree-like network of networks [4].

For a given function $\mathfrak{F}(x)$, we can solve Eqs.~(\ref{psia}) and (\ref{sia}) to get the giant cluster and the critical point. To show how to analysis the infinite generation, we also take the case used in the main text as an example, \emph{i.e.}, a two-layer network with the same degree distribution in each layer.

\subsubsection{Two-layer ER networks}

\begin{figure}
	\centering
	\scalebox{1.2}{\includegraphics{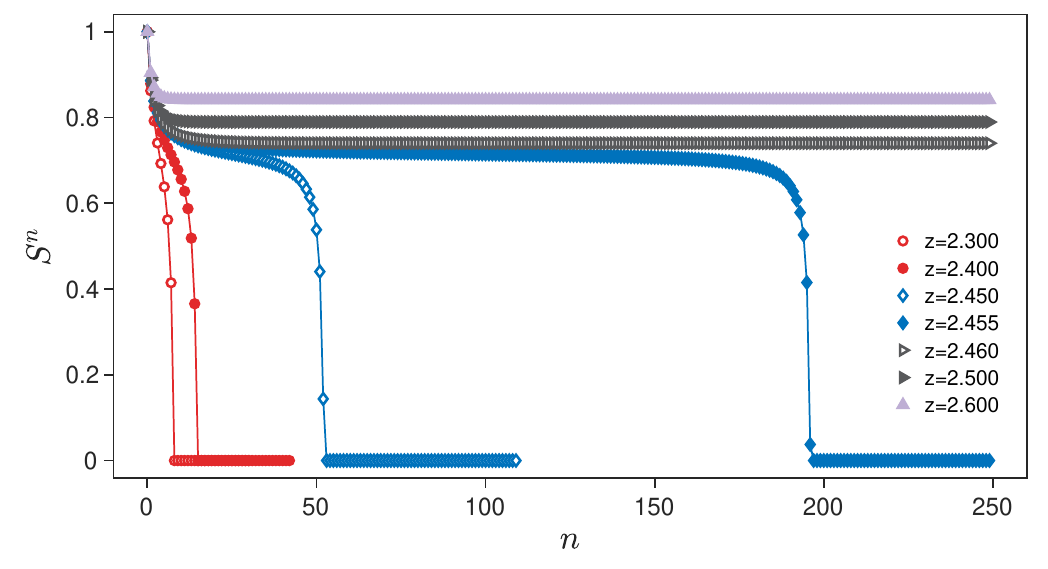}}
	\caption{The fixed points for Eqs.~(\ref{gns1s}) and (\ref{gns2s}). The curves are the numerical solutions of two-layer ER networks, for which the two layers have the same $\mathfrak{F}_(x)$ Eq.~(\ref{serp}).}
	\label{fig3s}
\end{figure}

Since the two layers have the same degree distribution, Eqs.(\ref{psia}) and (\ref{sia}) reduce to
\begin{eqnarray}
\psi &=& S^2, \\
S  &=& \mathfrak{F}(S). \label{sfs}
\end{eqnarray}
Here, $S^0=1$ is used. It is easy to know that the critical point is determined by Eq.~(\ref{sfs}). Together with Eq.~(\ref{serp}), we have
\begin{equation}
S = 1-e^{-zS^2}.
\end{equation}
Below a critical point $z_c^\infty\approx2.455$, this equation has only the trivial solution $S=0$ because the ``inverted Bell curve'' on the right hand side (rhs) is under the linear left hand side (lhs) for any $S>0$. At the critical point $z_c^{\infty}$, the lhs and rhs touch in one point. Denoting the rhs as $f(S)$, the touching point $S_c$ satisfies $f(S_c)=S_c$ and $f^\prime(S_c)=1$, which allows us to find its form
\begin{equation}
S_c=\frac{1}{2}\left(1+\sqrt{1-\frac{2}{z_c^\infty}}\right)\approx 0.7153.
\end{equation}
The other existing solution $S_c=(1-\sqrt{1-2/z_c^\infty})/2\approx 0.2847$, is smaller than the corresponding critical point $x_c=1/z_c^\infty\approx0.4073$ of function $\mathfrak{F}(x)$. Further iterations will thus turn it to $0$, which indicates that this smaller solution is not a real fixed point. The critical order parameter that corresponds to the found $S_c$ is $\psi_c^\infty=(S_c)^2>0$. The giant cluster size thus undergoes a discontinuous phase transition at $z_c^{\infty}$. Above the critical point, the fixed point $S$ further grows with $z$ as shown in Fig.~\ref{fig3s}, corresponding to the percolating state.

\subsubsection{Two-layer SF networks}

\begin{figure}
	\centering
	\scalebox{0.6}{\includegraphics{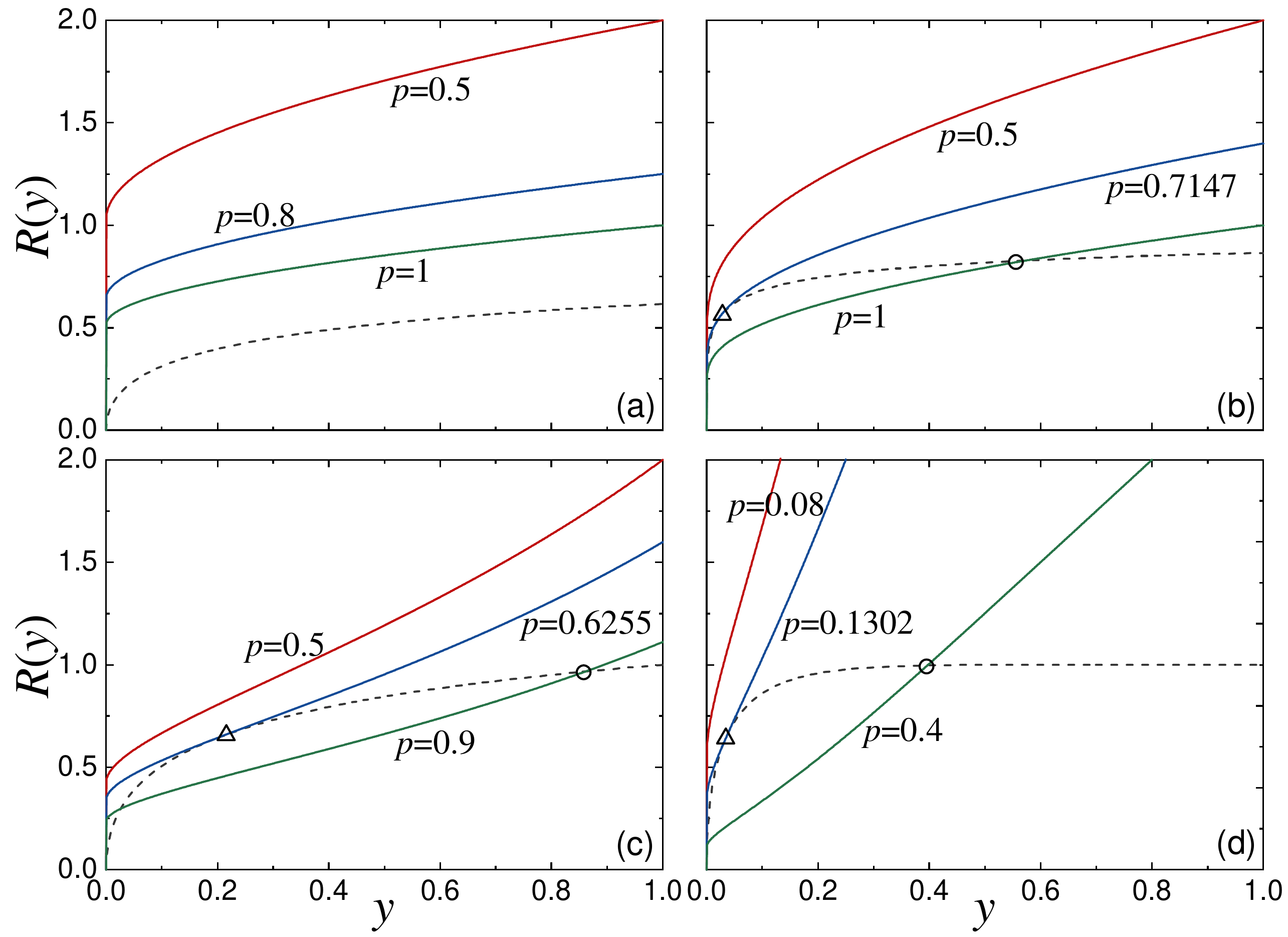}}
	\caption{A graphical representation of the numerical solution of Eqs.~(\ref{r1}) and (\ref{r2}). To obtain these curves, we use the maximum degree $K=10^5$. The dashed and solid lines represent functions $R_1(y)$ and $R_2(y)$, respectively. The triangles and circles mark the critical points and the non-trivial solutions, respectively. (a) $m=1$ and $\gamma=2.5$. (b) $m=1$ and $\gamma=2.1$. (c) $m=2$ and $\gamma=2.5$. (d) $m=8$ and $\gamma=2.5$.}
	\label{fig4s}
\end{figure}

Now, we study the solution of Eqs.~(\ref{sf1}) and (\ref{sf2}) for infinite generation, for which $x=\mathfrak{F}(px)$ and $\psi=x^2$. For qualitative analysis, supposing $p$ is fixed, and letting $y\equiv pRx$ in Eqs.~(\ref{sf1}) and (\ref{sf2}), we have
\begin{eqnarray}
R&=&1-G_1(1-y)\equiv R_1(y), \label{r1} \\
R&=&\frac{1}{p}\frac{y}{1-G_0(1-y)}\equiv R_2(y).  \label{r2}
\end{eqnarray}
The solution of these two equations can be represented by the cross points of functions $R_1(y)$ and $R_2(y)$ as shown in Fig.~\ref{fig4s}. When $y\to 0$, functions $R_1(y)$ and $R_2(y)$ behave as
\begin{eqnarray}
R_1(0)=0 &,& ~~\left.\frac{d R_1(y)}{d y}\right\rvert_{y=0} = G_1^\prime(1)\to\infty,\\
R_2(0)=0 &,& ~~\left.\frac{d R_2(y)}{d y}\right\rvert_{y=0} \sim \frac{1}{p}\frac{1}{1-G_0(1)}\to\infty.
\end{eqnarray}
It is not hard to find that $dR_2(y)/dy$ is a higher order infinity, thus $R_2(y)$ is larger than $R_1(y)$ in the area around $y\to 0^+$. In addition, for $y\to 1^-$, $R_1(y)$ and $R_2(y)$ behave as
\begin{eqnarray}
R_1(1)=1-\frac{p_1}{z}&,& ~~ \left.\frac{dR_1(y)}{dy}\right\rvert_{y=1}= G_1^\prime(0)=\frac{2p_2}{z},\\
R_2(1)=\frac{1}{p}&,& ~~ \left.\frac{dR_2(y)}{d y}\right\rvert_{y=1}=\frac{1-G_0^\prime(0)}{p}=\frac{1-p_1}{p}.
\end{eqnarray}
Here $p_1=c\approx(\gamma-1)m^{\gamma-1}$ (for $m\leq1$) is the fraction of the nodes with degree $1$, $p_2=c2^{-\gamma}\approx(\gamma-1)m^{\gamma-1}2^{-\gamma}$ (for $m\leq2$) is that for degree $2$, and $z\approx m(\gamma-1)/(\gamma-2)$ is the average degree.

When $m=1$, $p_1/z\approx\gamma-2$, even if $p=1$, $R_2(y)$ is larger than $R_1(y)$ when $y\to 1^-$. For large $\gamma$, such as $\gamma=2.5$ used in Fig.~\ref{fig4s}(a), there is no crossing point, meaning that no percolation transition exists for $m=1$. For comparison, Fig.\ref{fig3s}(b) shows non-trivial cross points for $\gamma=2.1$.

When $m>1$ (\emph{i.e.}, $p_1=0$), as shown in Fig.~\ref{fig4s}(c)-(d), the non-trivial cross points of functions $R_1(y)$ and $R_2(y)$ shift to the right as $p$ grows. The point of tangent with non-zero $R(y)$ and $y$ indicates that the percolation transition is discontinuous. In addition, as $m$ increases, the divergence of $dR_1(y)/dy$ becomes faster when $y\to 0^+$. The tangent point thus tends to the point $(0,0)$, meaning that a continuous percolation transition can also be found for infinite generation when $m\to\infty$ (average degree is then, naturally, also very large).

\section{Algorithm}
For simulations, our model can be implemented directly as the rules of the model, \emph{i.e.}, performing the percolation process iteratively, whose time complexity is $n\mathcal{O}(N)$ for generation $n$ and network size $N$. To be more efficient, the simulations in this paper are realized by a Leath-like method [5]. Instead of searching the whole network repeatedly, we evolve nodes (actually with some of its neighbors) one by one to the generation we want to study.

In our algorithm, each node has two variables, generation $n_i$ and root $r_i$, labeling which generation and cluster node $i$ belongs to. Here, $n_i$ must be non-negative, meaning that node $i$, as well as some of its neighbors, are in a cluster of generation $n_i$. In general, $r_i\geq 0$ is the root node of node $i$ (nodes are numbered from $0$ to $N-1$). If $-N\leq r_i<0$, it means that node $i$ is the root of the corresponding cluster and $-r_i$ is the number of nodes in this cluster. Based on this, the root node $root(i)$ of a cluster can be found from any node $i$ in this cluster. One possible way to do this (pseudo code) is as follows [6]:
\begin{eqnarray}
&&\textbf{root}~(i):   \nonumber  \\
&1&~~~~~~~~~j=i   \nonumber  \\
&2&~~~~~~~~~\textbf{while}~~~r_i\geq 0   \nonumber \\
&3&~~~~~~~~~~~~~~~\textbf{do}   \nonumber \\
&4&~~~~~~~~~~~~~~~~~~~~~r_j=r_i    \nonumber \\
&5&~~~~~~~~~~~~~~~~~~~~~j=i    \nonumber \\
&6&~~~~~~~~~~~~~~~~~~~~~i=r_i    \nonumber  \\
&7&~~~~~~~~~~~~~~~\textbf{end}   \nonumber  \\
&8&~~~~~~~~~\textbf{return}~~~i  \nonumber
\end{eqnarray}
Here, lines $4$ and $5$ are used to compress the path from node $j$ to the cluster root, which do nothing to this function $root(i)$ itself, but facilitate further searching.

Assuming that we want to obtain the percolation configuration of generation $n$, the main steps of our algorithm are as follows:
\begin{enumerate}
  \item Assigning each node a generation $n_i=0$ and a root $r_i=-N-1$, which means they have not yet been searched.
  \item For node $i$ with $n_i<n$, do the following steps:
      \begin{description}
        \item[a] Determining the network layer $l$ used in generation $n_i+1$ according to the model setting.
        \item[b] Finding all the nodes (labeled $j$ for convenience) connecting to node $i$ by layer $l$ with $n_j=n_i$ and $root(j)=root(i)$.
        \item[c] If the number of the found nodes $N_f=-r_{root(i)}$, let $n_j=n$ for all these nodes including node $i$; Otherwise, let $r_{root(i)}=-N_f$, and $n_j=n_j+1$ for all these nodes including node $i$.
        \item[d] If the new $n_i<n$, do steps (a)-(c) again.
      \end{description}
  \item Do step $2$ for all the nodes.
\end{enumerate}
Obviously, this algorithm is more effective than the one that just implementing the model rules directly, since the search times for most nodes are less than $n$ in generation $n$. Specifically, one can simply use the depth-first or breadth-first searching to implement step $2$ of this algorithm. To pursue a more effective way, other typical algorithms in percolation model can also be used to realize step $2$, which could depend on the measurement we are interested in. In addition, this type of algorithm has already been used in the percolation on the so-called interdependent networks[7].

\section{The largest cluster at the critical points}
\begin{figure}
	\centering
	\scalebox{0.55}{\includegraphics{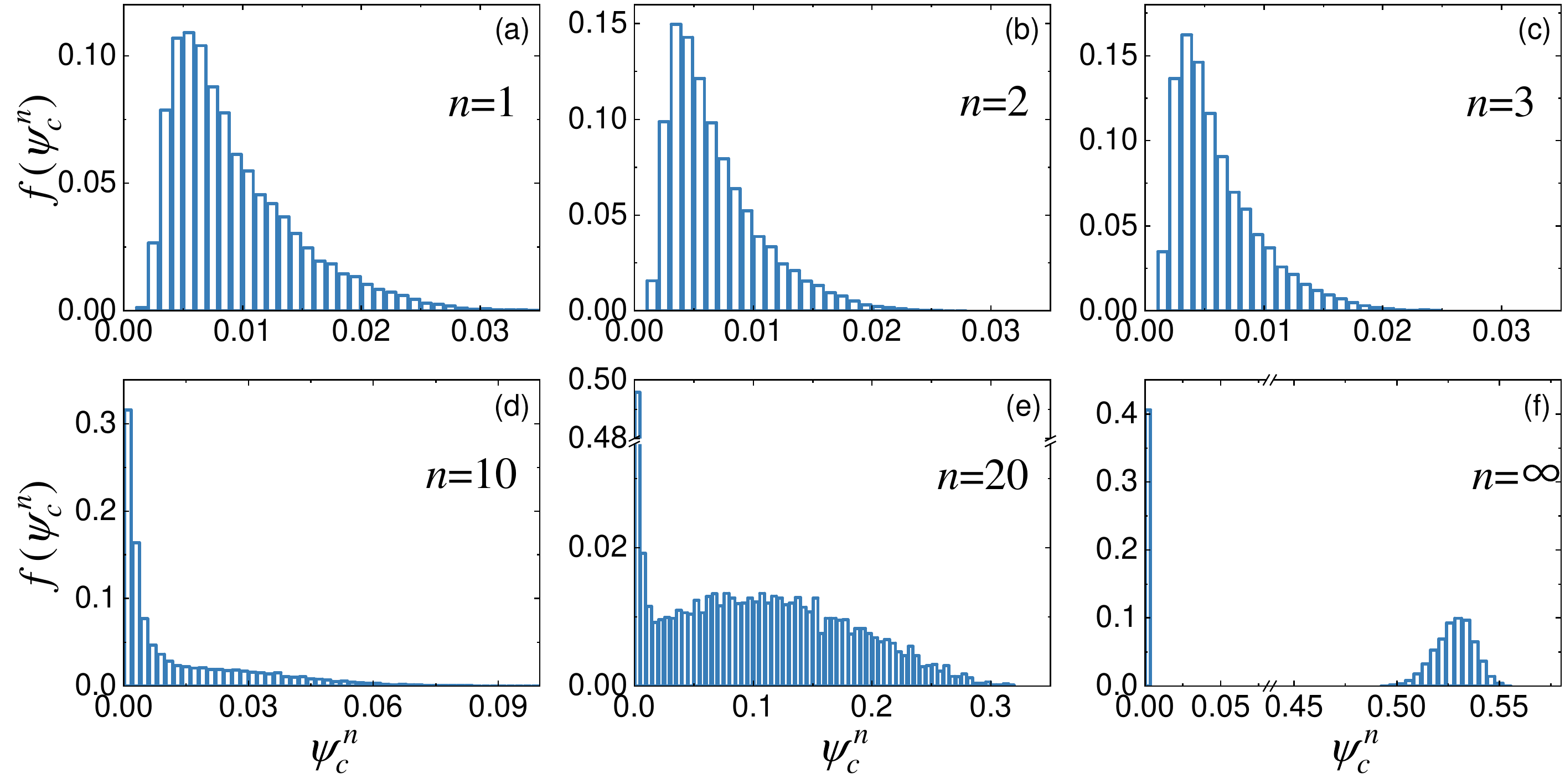}}
	\caption{The distribution of the order parameter $\psi_c^\infty$ at the critical point. The size of the network used in the simulation is $N=2^{20}$.}
	\label{fig5s}
\end{figure}

As the generation increases, the history-dependent percolation becomes sharper and sharper. This increases the finite-size effect around the critical point, which makes the simulation results appear to deviate from the finite-size scaling of a continuous phase transition. For a better understanding of this finite-size effect, we show the distribution of the order parameter $\psi_c^n$ in individual model realizations in Fig.~\ref{fig5s}.

For the first several generations (\emph{i.e.}, $n$ is not too large, see for examples Fig.~\ref{fig5s}(a)--(d)), the values of $\psi_c^n$ congregate in a small region. In these cases, we may not be able to distinguish the non-percolating and percolating realizations, since they all give a $\psi_c^n$ close to zero. In spite of this, the corresponding finite-size scaling still fit well with the theoretical analysis even for small network sizes (see Fig.2(b) in main text). However, with the increasing of $n$, the percolation transition becomes sharper and sharper, which produces excessively high percolation rates at the critical point for small systems. The results are the distribution of $\psi_c^n$ becomes broader and strongly bimodal for late generations ($n\gtrsim 20$), and the value obtained by averaging over these $\psi_c^n$ become larger than expectations. Therefore, the deviation from the finite-size scaling for large $n$ is caused by the consideration of these excessively percolating realizations.

When $n=\infty$ where the discontinuous transition exists, although theoretically the order parameter $\psi_c^\infty$ should be non-zero at the critical point, a finite system at (even above) the critical point can also fail to percolate and thus the bimodal distribution of $\psi_c^\infty$ is still present. The prominent bimodal distribution of $\psi_c^\infty$ (see Fig.~\ref{fig5s}(f)) allows the non-percolating and percolating realizations to be distinguished: $\psi_c^\infty$ close to zero for the non-percolating realizations and larger values for percolating realizations. If the evaluation of simulations is limited to the percolating realizations, the expected behavior of $\psi_c^\infty$ is again recovered (see Fig.2(d) in main text).

\section{Social network}
\begin{figure}
\centering
\scalebox{1.3}{\includegraphics{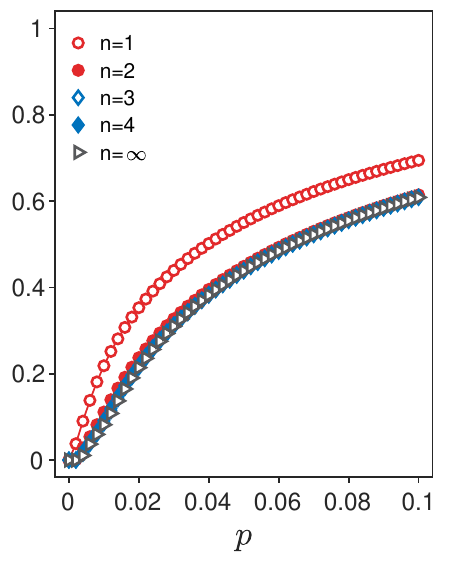}}
\caption{History-dependent percolation on social network composed of users who are active on both Twitter and FriendFeed. The average degrees are $z\approx110$ and $70$ in the two network layers, respectively. Here $p$ is the link occupation probability.}
\label{fig6s}
\end{figure}

In Fig.~\ref{fig6s}, we apply our model on a social network composed of users who are active on both Twitter and FriendFeed. Among the $150,684$ common users of the two networks, there are $8,308,326$ and $5,270,665$ links in the Twitter and FriendFeed layer, respectively.

\section{Brain network}
\begin{figure}
\centering
\scalebox{1.3}{\includegraphics{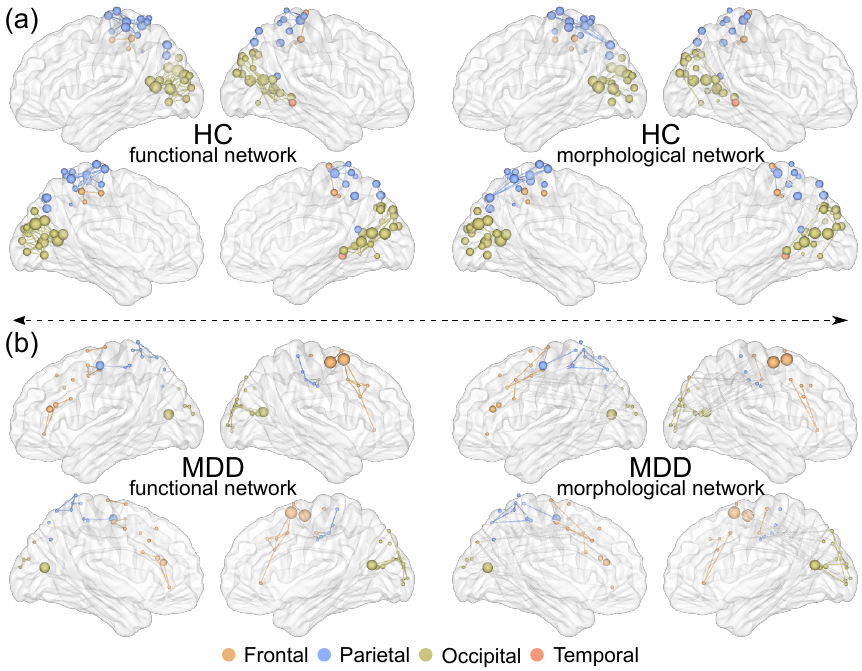}}
\caption{The visualization of human brain two layers networks at degree $z=6$ and $1024$ parcellation templates when $p$ is slightly larger than the corresponding $p_c$. (a) The HC participant's bilayer brain network. (b) The MDD participant's bilayer brain network.}
\label{fig7s}
\end{figure}

In Fig.~\ref{fig7s}, we give the pattern of the remaining nodes and links (the giant cluster of infinite generation) of HC and MDD participants when $p$ is slightly larger than $p_c$. We can find that for $z=6$ the remaining nodes and links of MDD participant is more sparse and dispersed, and mainly located in frontal, parietal, and occipital lobes, and no obvious community structure formed.

\section{MRI data and processing}
We selected a healthy control (HC) participant (subject ID:206525) and obtain the corresponding resting-state functional magnetic resonance imaging (R-fMRI) data and T1-weighted data from the ``S1200" release of the Human Connectome Project data (1U54MH091657) [8]. We similarly obtained the R-fMRI and T1-weighted data of one major depressive disorder (MDD) participant from our own dataset for comparison. The entire processing of the R-fMRI data is conducted using the Statistical Parametric Mapping (SPM12, version r7219) [9] and our own MATLAB codes. We processed the T1-weighted data with the Computational Anatomy Toolbox (CAT12, version r1278) [10].

\emph{R-fMRI data processing}. In the current study, our R-fMRI data processing includes the following steps: 1) Removing the volumes of the first $10$ seconds; 2) Realignment of all volumes to the first volume; 3) Mean-based intensity normalization; 4) Spatial normalization to the MNI template with EPI ($2\times 2\times 2 mm$ voxel size); 5) Linear detrending of retain mean and bandpass filtering between $0.043$ and $0.087 Hz$ using the Butterworth filter; 6) Denoising: 24HMP$+$8Phys$+$4GSR$+$Spikereg [11]; 7) Each fMRI voxel value is weighted by the gray matter probability.

\emph{T1-weight data processing}. The raw MRI data were checked manually to ensure no obvious artifacts. We use the CAT12 toolbox to perform the voxel-based morphometry analysis and the T1-weighted image is segmented into gray matter (GM), white matter and cerebrospinal fluid. The resulting GM images are normalized to the MNI space and undergo nonlinear modulation. Finally, the GM volume images for each participant are obtained.

\section*{References}
\begin{enumerate}
  \item William Aiello, Fan Chung, and Linyuan Lu. A random graph model for power law graphs. Experimental Mathematics, 10(1):53–66, 2001.
  \item Reuven Cohen, Daniel ben Avraham, and Shlomo Havlin. Percolation critical exponents in scale-free networks. Phys. Rev. E, 66:036113, 2002.
  \item D.-S. Lee, K.-I. Goh, B. Kahng, and D. Kim. Evolution of scale-free random graphs: Potts model formulation. Nucl. Phys. B, 696(3):351 – 380, 2004.
  \item J. Gao, S. V. Buldyrev, S. Havlin, and H. E. Stanley. Robustness of a network of networks. Phys. Rev. Lett., 107:195701, 2011.
  \item P. L. Leath. Cluster size and boundary distribution near percolation threshold. Phys. Rev. B, 14:5046–5055, 1976.
  \item M. E. J. Newman and R. M. Ziff . Fast monte carlo algorithm for site or bond percolation. Phys. Rev. E, 64:016706, 2001.
  \item P. Grassberger. Percolation transitions in the survival of interdependent agents on multiplex networks, catastrophic cascades, and solid-on-solid surface growth. Phys. Rev. E, 91:062806, 2015.
  \item D. C. Van Essen, S. M. Smith, D. M. Barch, T. E. J. Behrens, E. Yacoub, and K. Ugurbil. The WU-minn human connectome project: An overview. NeuroImage, 80:62–79, 2013.
  \item http: // www.fil.ion.ucl.ac.uk / spm / software / spm12.
  \item http: // www.neuro.uni-jena.de / cat12.
  \item L. Parkes, B. Fulcher, M. Y\"{u}cel, and A. Fornito. An evaluation of the e ﬃ cacy, reliability, and sensitivity of motion correction strategies for resting-state functional mri. NeuroImage, 171:415–436, 2018.
\end{enumerate}

\end{document}